\documentclass[preprint,12pt]{elsarticle}

\usepackage{amssymb}
\usepackage{amsmath}

\usepackage{amsmath,amsfonts}
\usepackage{url}
\usepackage{graphicx}
\usepackage{tabularx,booktabs,diagbox}
\usepackage{multicol}
\usepackage{multirow}
\usepackage{soul}
\usepackage{titlesec}
\usepackage{subfig}

\usepackage[dvipsnames]{xcolor}

\journal{Nuclear Physics B}

\begin{document}

\begin{frontmatter}



\title{A comparison between black-, grey- and white-box modeling for the bidirectional Raman amplifier optimization} 


\author[1]{Metodi P. Yankov} 
\author[1]{Mehran Soltani}
\author[2]{Andrea Carena}
\author[1]{Darko Zibar}
\author[1]{Francesco Da Ros}

\affiliation[1]{organization={Department of Electrical and Photonic Engineering, Technical University of Denmark},
            addressline={Oersteds Plads, Building 343}, 
            city={Kgs. Lyngby},
            postcode={2800}, 
            country={Denmark}}

\affiliation[2]{organization={Dipartimento di Elettronica e Telecomunicazioni (DET), Politecnico di Torino},
            addressline={Corso Duca degli Abruzzi, 24}, 
            city={Torino},
            postcode={10129 }, 
            country={Italy}}
\ead{fdro@dtu.dk}

\begin{abstract}
Designing and optimizing optical amplifiers to maximize system performance is becoming increasingly important as optical communication systems strive to increase throughput.  Offline optimization of optical amplifiers relies on models ranging from white-box models deeply rooted in physics to black-box data-driven and physics-agnostic models. Here, we compare the capabilities of white-, grey- and black-box models on the challenging test case of optimizing a bidirectional distributed Raman amplifier to achieve a target frequency-distance signal power profile. We show that any of the studied methods can achieve similar frequency and distance flatness of between 1 and 3.6 dB (depending on the definition of flatness) over the C-band in an 80-km span. Then, we discuss the models' applicability, advantages, and drawbacks based on the target application scenario, in particular in terms of flexibility, optimization speed, and access to training data.
\end{abstract}

\begin{graphicalabstract}
\end{graphicalabstract}

\begin{highlights}
\item Research highlight 1
\item Research highlight 2
\end{highlights}


\end{frontmatter}



\section{Introduction}
The behavior of optical amplifiers directly affects the signal performance in optical communication systems~\cite{rapp2021optical} and their impact grows as communication systems get extended to multi-band~\cite{rapp2021optical} and/or space-division multiplexing~\cite{puttnam2021space} operation. 
Therefore, modeling and more importantly optimizing the amplifier response is highly desirable, and considerable research effort has been dedicated to the topic~\cite{Yankov:23,jones2023spectral,yang2023experimental,minakhmetov2023digital}. Substantial earlier work on modeling optical amplifiers has been mainly focused on describing the physical processes leading to optical amplification \cite{Agrawal, Saleh1990,PelouchJLT16,islam2002raman}, referred to as white-box models. With such models, offline optimization of the gain profile of the amplifiers can be carried out. This process is generally cumbersome due to the often complicated differential equations involved in predicting the amplifier behavior. Furthermore, the physical parameters involved, e.g. gain-medium properties such as emission spectra \cite{Saleh1990}, gain coefficient spectra (Raman)~\cite{PelouchJLT16,islam2002raman}, etc., are not always easy to characterize. Model fitting is therefore applied~\cite{borraccini2023gain,hafermann2023preemphasis,jones2023spectral,eldahrawy2024parameter} but inaccuracies between the fitted model and the physical amplifier directly impair the optimization. Additionally, iterative optimization may suffer from a significant computational complexity if such white-box models are non-differentiable.

Alternatively, online optimization relying on gradient-free methods can be applied directly to the physical device. These mainly rely on the use of evolutionary algorithms, e.g. genetic algorithms \cite{Neto:07,li2014design,borraccini2022cognitive}. While they do not require a model, the optimization of live systems is challenging and time-consuming due to the requirement of feedback from the physical system, system downtime during the optimization process, and the related measurement uncertainties, which might mislead the optimization.

More recently, black-box models, based e.g. on neural networks (NNs), have been proposed \cite{zhou2006robust, chen2018optimal, deMoura:20, 9333241,ye2020experimental,mineto2021performance,da2021optimization,donodin2023neural,minakhmetov2023digital}. These models can be trained on either synthetic or experimental data and can directly implement inverse models - given the target amplifier response they provide the amplifier parameters - enabling nearly real-time reconfigurability once the models are trained. However, a relatively large training dataset is normally required, especially if generalizable models are desired, e.g. in terms of multiple Erbium-doped fiber amplifier (EDFA) units~\cite{yankov2021power} or Raman amplifier relying on different fiber types~\cite{de2022fiber}. 

Finally, the use of physics-informed NN models has gained increasing interest following their first introduction in~\cite{raissi2019physics}. These models combine the data-driven approach of black-box NN with the knowledge from the underlying physics, generally included through an ad-hoc loss function which includes terms from both the data-driven and the physics-informed sides. Such an approach has also seen preliminary applications to nonlinear propagation in optical fibers~\cite{jiang2023predicting} more specifically to Raman amplifiers~\cite{mei2024power,zhang2024,liu2022physics}.

All of the above approaches, from the white-box physical modeling to the black-box NN-based models, have advantages and drawbacks depending on the target application~\cite{da2023modeling}. Additionally, a systematic comparison of their performance on the same test case has yet to be reported.

In this work, we compare four optimization methods spanning the categories just described. Their performance is studied on the particularly challenging test case of a bi-directional distributed Raman amplifier with first and second-order pumping where the wavelength division multiplexed (WDM) signal power evolution over frequency and fiber distance is jointly optimized, aiming for maximizing flatness over both dimensions. Then, the applicability of each method to different optimization scenarios is discussed.
In Section~\ref{sec:Method}, the considered optimizers are briefly described. The test scenario is defined in Section~\ref{sec:Setup}. The methods' performance and applicability are reported in Sections~\ref{sec:Performance} and~\ref{sec:Applicability}, respectively. Section~\ref{sec:Conclusions} summarizes the main findings.

\section{Raman optimizers} 
\label{sec:Method}
The signal and pump power evolution through a Raman amplifier can be numerically calculated by solving a system of nonlinear partial differential equations \cite{9721641}:

\begin{align}
\label{eq:Raman}
 \frac{\partial P_m(f, z)}{\partial z} & = \pm 2\alpha_m(f) P_m(f, z)  +\sum_{f_n \in \mathcal{F}}\frac{g_R(f_n-f_m)}{A_{eff}}{P(f_m, z) \cdot P(f_n, z)},
\end{align} 

where the indexes $m$ refers to the current carrier (pump or signal), the index $n$ indexes the co- and counter-propagating carriers (positive and negative sign on the first right-hand side term) and ${\mathcal{F}} = {\mathcal{F}_p} \cup {\mathcal{F}_s} $ is the set of active carriers including the set of pumps ${\mathcal{F}_p}$ and the set of signals ${\mathcal{F}_s}$. $P(f, z)$ is the carrier power at frequency $f$ and distance $z$, $g_R(f_n-f_m)$ is the Raman gain coefficient for a given offset between the frequencies $f_n$ and $f_m$ (here, assumed independent of the reference wavelength and constant over the C-band), $\alpha(f_n)$ is the fiber loss at frequency $f_n$ and $A_{eff}$ is the fiber effective area.
This model does not allow for analytical optimization, as it lacks a closed-form analytical solution. Therefore different approaches have been proposed and are briefly described below.

\textit{Optimizers based on NNs.}
Neural networks can approximate complex nonlinear relations which may be challenging to express analytically with tractable models. That is indeed the case for the relation between a Raman amplifier response (signal 2D evolution in frequency and fiber distance) and its input parameters (pump powers and wavelengths).  For the 2D evolution, we rely on the convolutional NN (CNN) proposed in \cite{Soltani:20}. The network takes in input the desired 2D signal evolution in frequency and distance and outputs the predicted set of pump powers and wavelengths expected to lead to such a target evolution, i.e. an inverse model. The CNN extracts the key features in the input space through a series of convolutional and pooling layers and makes the final prediction through a fully connected regression layer. Such an offline model can be trained using synthetic \cite{Soltani:20} or experimental \cite{Soltani_OE} data. Once the model is trained, the inference is nearly instantaneous as it simply relies on a sequence of convolution operations, matrix multiplications, and nonlinear activation functions (ReLU in our specific implementation). As the model is trained on a specific dataset, the CNN is constrained to the configurations (number of pumps, number of signal channels, fiber properties, etc.) that have been used to generate such a dataset, and extrapolation beyond that typically leads to poor accuracy. In this work, a training dataset of 3500 samples is generated using the Raman solver provided by the GNPy library \cite{Gnpy}. The topology of the CNN is optimized to 3 convolutional layers with 32 filters each with a kernel size of 3x3 in each layer followed by max-pooling, and finally two fully-connected regression layers with 40 hidden nodes. The spatial resolution for the network input is fixed to 500~m. More details can be found in \cite{Soltani:20}.

\textit{Gradient-free optimizers.}
Gradient-free optimization algorithms can be applied to numerical and analytical models which are non-differentiable and more importantly to experimental setups where a gradient cannot be explicitly calculated. Among gradient-free optimizers, evolutionary algorithms, such as differential evolution (DE), rely on the evolution of a population of solutions that approach the desired target by iteratively penalizing the individuals of the population that achieve high values of the cost function. Here, we apply a DE algorithm to optimize the pump parameters providing target 2D power profile which are calculated by the numerical Raman solver. The algorithm uses the optimized parameters of \cite{9721641} for our specific test scenario: a crossover probability, mutation factor, and population size of 0.5, 0.8, and 30, respectively.

\textit{Neural network + gradient-free optimization.}
As gradient-free optimizers operate iteratively, their convergence (if any) can be highly dependent on the initial conditions provided to the optimizer. In order to speed up the convergence and the quality of the final solution, an approximate initial condition can be provided. In the proposal of \cite{9721641}, the initial population provided to the DE is constructed by choosing pump powers and wavelengths within a small range of the prediction provided by the CNN-based model. This choice focuses the DE to a promising subset of the overall search space, thus significantly reducing the number of iterations needed for convergence but at the cost of requiring a pre-trained CNN model and assuming that the CNN prediction is of a sufficiently high accuracy. However, DE can then extrapolate beyond the CNN training data set without the accuracy penalty affecting the CNN. A similar approach has been applied to the more recent work of ~\cite{gao2024fast}, where the gradient-free optimizer can be used to correct for modeling inaccuracies or variations over time. This combination of a white-box model (DE) with a black-box (NN)-based initialization can be considered a grey-box model, similar to physics-inspired NN models~\cite{raissi2019physics,mei2024power,jiang2023predicting,zhang2024}. 

\textit{Gradient-based optimizers.}
The signal and pump power evolution through a Raman amplifier can be numerically calculated by using finite difference methods to solve (\ref{eq:Raman}), i.e. in a white-box approach. 
Such a system of equations can be mostly expressed in a frequency-differentiable form with the key exception of the Raman gain efficiency which is normally either calculated by numerically computing an integral or defined by a look-up table based on experimental measurements. Neither option is differentiable. In order to overcome this limitation, in \cite{9244561, Yankov:23} it was proposed to fit a differentiable function to the Raman gain efficiency. If a sufficiently accurate fit of the Raman gain is achieved (e.g. with an overfitted neural network \cite{Yankov:23}), the accuracy of the solver solution is not compromised and the full system of equations becomes fully differentiable. Gradient descent (GD)-based optimization can therefore be applied allowing differentiability over pump, signal, and fiber parameters. Remark that the use of a GD-based optimizer for backward or bidirectional Raman amplifiers does not require the use of time-consuming iterative algorithms (e.g. shooting algorithms) that are normally required to tackle the boundary-value problem. GD-based optimizers simply fine-tune the powers at the input of the fiber for all pumps. The input pump powers, then, uniquely determine the powers at the output of the fiber for the counter-propagating pumps~\cite{9244561}. Therefore, the ability to use GD-based optimization over the system of (\ref{eq:Raman}) can provide a powerful alternative to physics-inspired NN models for specific applications. 

\section{Setup}
\label{sec:Setup}

\begin{figure}[h]
\centering\includegraphics[width=12cm]{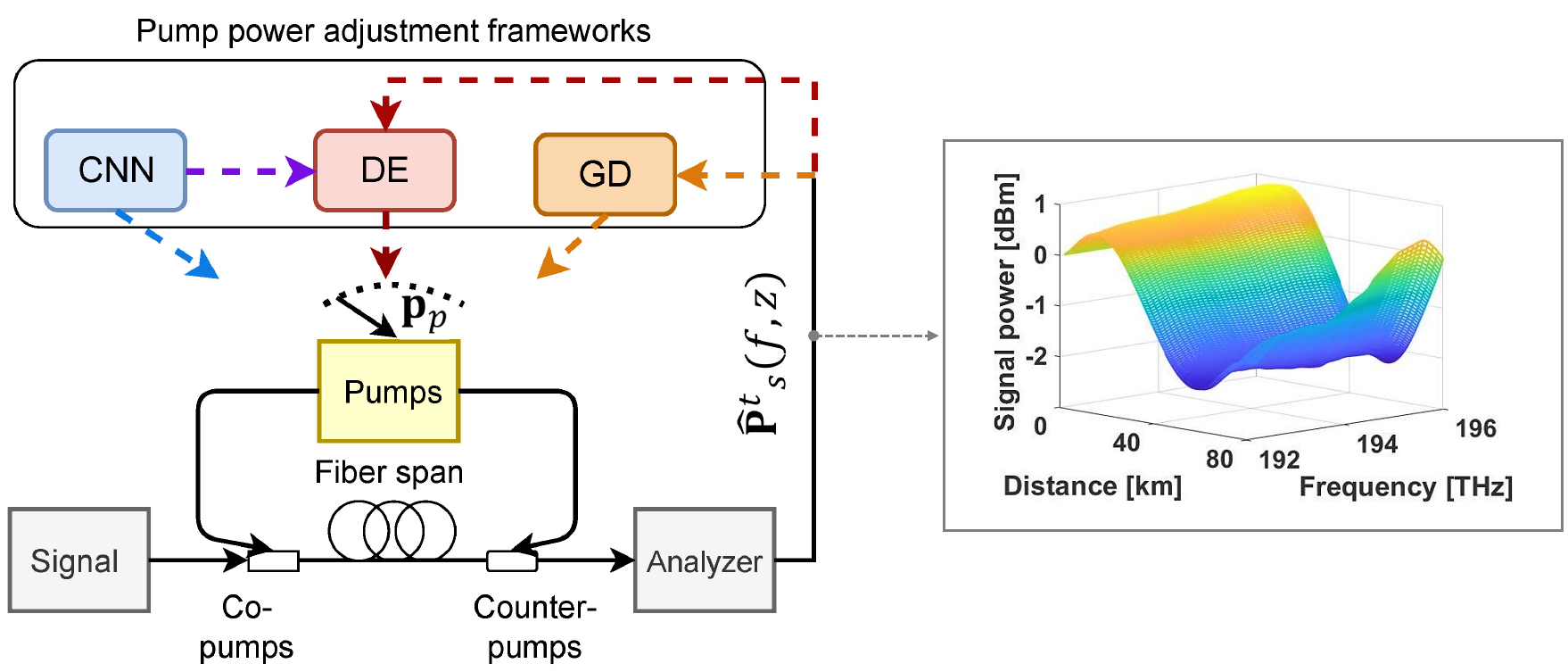}
\caption{The proposed distributed Raman amplifier setup with different pump power adjustment frameworks. The frameworks are discriminated with different colors (CNN: Blue, DE: Red, CNN+DE: Purple+Red, GD: Orange). The inset shows an example of a simulated 2D power profile for a given pump configuration and flat input power channel load.}
\label{fig:Setup}
\end{figure}

The considered setup for the distributed Raman amplifier is shown in Fig.~\ref{fig:Setup} and is based on $L=80$ km of a standard single-mode fiber (SSMF) with 0.2 dB/km, 0.25 dB/km and 0.32 dB/km of attenuation for 15xx-nm signal and 14xx-nm and 13xx-nm pumps, respectively. Effective area, nonlinear coefficient and peak Raman gain coefficient are set to 80~$\mu m^2$, 1.26~1/W/km, 0.3841~1/W/km. Raman gain efficiency spectrum of an SSMF is assumed~\cite{Gnpy}.

The optimization is carried out for 40 100-GHz spaced signal channels at a launch power of 0~dBm/channel spanning the C-band (192 THz to 196 THz). Bidirectional pumping is assumed with three first-order pumps and one second-order pump per propagation direction (eight pumps in total). The pump wavelengths and pump power constraints are defined in Table~\ref{tb:pumpRanges}. 

\begin{table}[h]
\centering
\caption{Raman pump wavelengths and power ranges.}

\label{tb:pumpRanges}

\begin{tabular}{c||c|c|c|c}

    Co-pumps & $p_1$ & $p_2$ & $p_3$ & $p_4$ \\ \hline\hline
    $\lambda$, nm & 1366 & 1425 & 1455 & 1475 \\\hline
     $[\textbf{p}_{LB}, \textbf{p}_{UB}]$, mW  & [200,1200]  & [5,150] & [5,150]& [5,150]\\ \hline
     $[\textbf{p}_{LB}, \textbf{p}_{UB}]$, dBm  & [23.0,30.8]  & [7.0,21.8] & [7.0,21.8]& [7.0,21.8]\\ \hline
    \\
    \hline
    Counter-pumps & $p_5$ & $p_6$ & $p_7$ & $p_8$ \\ \hline
    $\lambda$ [nm]  & 1366 & 1425 & 1455 & 1475 \\\hline
    $[\textbf{p}_{LB}, \textbf{p}_{UB}]$, mW  & [200,1200] & [5,150]& [5,150]& [5,150] \\ 
    \hline
    $[\textbf{p}_{LB}, \textbf{p}_{UB}]$, dBm  & [23.0,30.8] & [7.0,21.8]& [7.0,21.8]& [7.0,21.8] \\ 
    \hline
\end{tabular}
\vspace{-0.3cm}
\label{tbl:pumps}
\end{table}

The four methods compared are: (i) the stand-alone black-box model based on a CNN, (ii) the gradient-free grey-box model based on the DE, (iii) the combined CNN+DE method, where the CNN provides the initial conditions to the DE-based optimizer, and (iv) the white-box model based on a differentiable implementation of eq.~(\ref{eq:Raman}) optimized through GD.
As mentioned, the CNN does not require feedback from the channel and provides a solution for the pump configuration `instantaneously'. The DE and GD algorithms are iterative optimization processes. At every iteration, the frequency-distance power profile $\hat{\mathbf{P}}_s(f, z)$ at the channel frequencies is calculated using (\ref{eq:Raman}). The new profile is used to update the pump configuration for the next iteration. Remark that in the case of DE, the calculated power profile can be replaced by an experimental measurement, e.g. through a frequency-swept optical time-domain reflectometer as in \cite{Soltani_OE}.

The chosen test case focuses on optimizing just the pump powers. However, all four methods also allow for optimization of the pump wavelengths, as well as fiber parameters or input signal power, at the cost of a more complex optimization for DE and GD-based methods and a larger dataset required for the CNN-based methods. By its inherent design, the GD-method applied to the differentiable implementation of (\ref{eq:Raman}) can effectively calculate derivative with respect to any signal, pump, and fiber parameter.

\section{Performance comparison}
\label{sec:Performance}
 A common scenario in multi-span transmission systems is to aim for flatness of the gain spectrum over distance and frequency. Depending on the application, flatness across the two dimensions is not of equal importance. It is therefore difficult to design a universal metric to benchmark the system and optimize the performance. Similarly, without such a metric, it is difficult to design a cost function for optimization. In the following, three heuristically chosen criteria and optimization strategies are proposed, aiming to cover a wide variety of applications. It is not the purpose of this paper to propose a universal strategy but to compare different optimization methods for the same target.

Without loss of generality to the target frequency-distance evolution definition, performance is estimated based on three main criteria, following~\cite{9721641}. All criteria relate to the frequency-distance excursions of the signal carriers ($f \in \mathcal{F}_s$) in dB scale. The first criterion is the maximum power excursion defined as
\begin{equation}
\label{eq:F0}
 J_0(\textbf{p}_{p}) = {\displaystyle\max_{f \in \mathcal{F}_s,z}(P(f,z|\textbf{p}_{p}))} -  {\displaystyle\min_{f \in \mathcal{F}_s,z}(P(f,z|\textbf{p}_{p}))},
\end{equation} 
where $P(f,z|\textbf{p}_{p})$ is the signal power at frequency $f$ and distance $z$ given the pump configuration $\textbf{p}_{p}$. The criterion $J_0$ is an ultimate goal and is the most difficult to achieve. Observe, that ultimate limits for $J_0$ are not theoretically known, but are interesting areas for future research to explore. In particular, the possibility for $J_0=0$ and its corresponding pump configuration would be of great theoretical and practical interest. \\
Second, the maximum power excursion over frequencies at any given distance is defined as
\begin{equation}
\label{eq:F1}
 J_1(\textbf{p}_{p}) =\smash{\displaystyle\max_{z}[\smash{\displaystyle\max_{f \in \mathcal{F}_s}(P(f,z|\textbf{p}_{p}))} - \smash{\displaystyle\min_{f \in \mathcal{F}_s}(P(f,z|\textbf{p}_{p}))}]}.
\end{equation}
Minimizing $J_1$ instead of $J_0$ is typically easier under a given pump constraint, as it allows for the channel powers to vary along the distance dimension, as long as the WDM signal maintains a compact frequency spectrum. \\
Third, the maximum deviation from a loss-compensated transmission (0~dB net gain) over frequency 
\begin{equation}
\label{eq:F2}
 J_2(\textbf{p}_{p}) = \smash{\displaystyle\max_{f \in \mathcal{F}_s}|P(f,L|\textbf{p}_{p}) - P(f,0|\textbf{p}_{p})}|.
\end{equation}
Minimizing $J_2$ is the easiest target and is the most practically relevant criterion, as it is directly related to the system requirement for additional frequency flattening filters and/or additional gain, e.g. through an EDFA~\cite{da2021optimization}. 

\begin{figure}[h]
\centering\includegraphics[width=7cm]{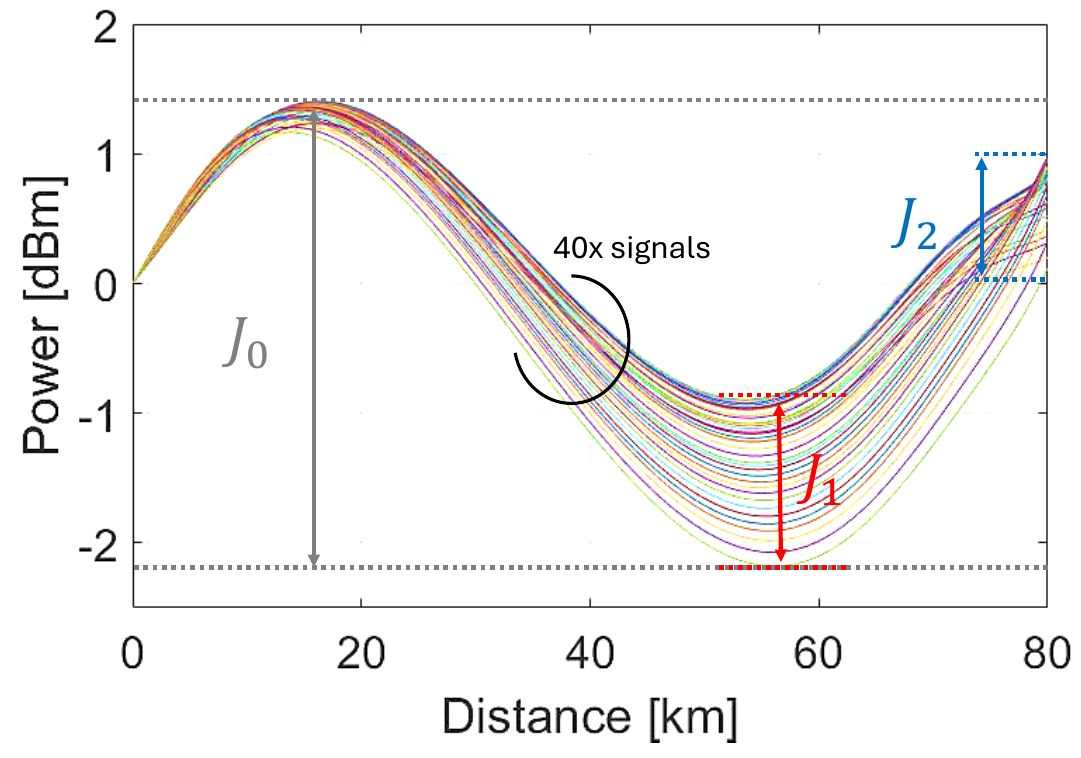}
\caption{Power evolution of the 40 signal channels over fiber distance showing the peak-peak estimation of the three loss criteria: maximum power excursion ($J_0$, grey), maximum power excursion over frequency ($J_1$, red), and deviation from loss-compensation ($J_2$, blue).}
\label{fig:loss}
\end{figure}

A visual representation of the three loss criteria is shown in Fig.~\ref{fig:loss}.

To provide a broader comparison of the four optimization methods, multiple application scenarios are considered by investigating different weighted cost functions $\textbf{m}^{(i)}$~\cite{9721641}. This choice aims at covering applications that give different weight/priority to the different criteria:
\begin{equation}
\begin{split}
     &\textbf{m}^{(0)} = \hskip0.7em J_0, \\
 &\textbf{m}^{(1)} = \frac{2}{3}J_0+\frac{1}{3}J_1, \\
 &\textbf{m}^{(2)} = \frac{2}{3}J_0+\frac{1}{6}J_1+\frac{1}{6}J_2.
\end{split}
\label{eq:loss}
\end{equation}

The loss functions of eq. (\ref{eq:loss}) are used for all optimization schemes, except for the stand-alone CNN, for which the criteria $J_0, J_1$, and $J_2$ cannot directly be incorporated in the training procedure as they are limited to the profiles available in the training set. Instead, the CNN is trained on the mean square error (MSE) over all distances and frequencies and is similarly producing a pump profile aiming at minimizing the MSE.\\
The cost function $\textbf{m}^{(0)}$ aims at providing a baseline by minimizing the difficult target.  The cost function $\textbf{m}^{(1)}$ aims at relaxing the requirement for ultimate flatness in favor of minimizing a more practically relevant and potentially easier to achieve metric: $J_1$. Similarly, $\textbf{m}^{(1)}$ further imposes the practical requirement that the gain spectrum is frequency flat. 
For a fair comparison, the performance in terms of $J_0$, $J_1$, and $J_2$ is evaluated based on the same solver applied to (\ref{eq:Raman}) for all modeling approaches, including the CNN-based. The performance w.r.t. each criterion for all optimization methods when each cost function is used is summarized in Table~\ref{tb:Performance}. For completeness, $J_1$ is given also in the case when $\textbf{m}^{(0)}$ is used, and $J_2$ is given in the cases when $\textbf{m}^{(0)}$ and $\textbf{m}^{(1)}$ are used, i.e. cases in which the specific criterion evaluated was not included in the overall cost function during optimization. These values may also be of interest in the case when a multi-purpose configuration is required. 

\begin{table*}[h]
\caption{Comparison of accuracy performance achieved by the different methods.}

\centering

\begin{tabular}{ m{1cm} || m{0.7cm} || m{0.7cm} | m{0.7cm} | m{0.7cm}|| m{0.7cm} | m{0.7cm}| m{0.7cm}|| m{0.7cm}| m{0.7cm}| m{0.7cm}}

     & \multicolumn{1}{c||}{CNN} & \multicolumn{3}{c||}{DE} & \multicolumn{3}{c||}{CNN+DE} & \multicolumn{3}{c}{GD}\\ 
model & \multicolumn{1}{c||}{only}  & $\textbf{m}^{(0)}$ & $\textbf{m}^{(1)}$ & $\textbf{m}^{(2)}$ & $\textbf{m}^{(0)}$ & $\textbf{m}^{(1)}$ & $\textbf{m}^{(2)}$ & $\textbf{m}^{(0)}$ & $\textbf{m}^{(1)}$ & $\textbf{m}^{(2)}$\\ \hline\hline
    $J_0$[dB] &3.58 & 2.82 & 3.04 & 3.11 & 2.81 & 2.97 & 3.06 & 2.61 & 2.79 & 2.86 \\ \hline
    $J_1$[dB] &1.48 & 1.63 & 0.82 & 0.96 & 1.80 & 0.88 & 0.90 & 1.98 & 0.77 & 1.03 \\ \hline
    $J_2$[dB] &0.97 & 2.28 & 2.86 & 1.18 & 1.14 & 1.20 & 0.65 & 2.50 & 2.50 & 1.11 \\ \hline
\end{tabular}
\label{tb:Performance}
\end{table*}

Generally, the different methods achieve similar performance in terms of $J_0$ to within $\pm 0.15$~dB, except the stand-alone CNN which is slightly penalized due to its dataset-restricted nature. The gradient-free stand-alone DE is slightly worse than its GD-based counterpart. The accuracy of DE is improved by initializing it using the CNN, in which case the overall best performance is achieved in terms of$J_1$ and $J_2$. The GD was initialized with the maximum allowed values from Table~\ref{tbl:pumps}. 

It is noted that even though the performance is similar, it is achieved with different pump configurations in all cases as shown in Table~\ref{tb:powerValues}, indicating the highly non-convex nature of the cost function. Initialization of all methods is thus important if more restrictive constraints on the pumps are to be imposed than the ones in Table~\ref{tbl:pumps}.

\begin{table}[t]
\caption{Predicted pump power values in dBm. DE (stand-alone and CNN+DE) and GD are trained for $\textbf{m}^{(2)}$, while the CNN still uses MSE as cost function.}
\centering
\label{tb:powerValues}
\begin{tabular}{c||c|c|c|c}
\centering
    \backslashbox[22mm]{pump}{model}
    & CNN & DE & CNN+DE & GD\\ \hline\hline
    $p_1$ [dBm] & 25.2 & 29.1 & 26.5 &  26.9\\ \hline
    $p_2$ [dBm] & 15.2 & 14.2 & 16.7 &  16.8\\ \hline
    $p_3$ [dBm] & 21.6 & 16.0 & 18.8 &  17.4\\ \hline
    $p_4$ [dBm] & 10.8 & 16.5 & 11.5 &  16.1\\ \hline
    $p_5$ [dBm] & 30.1 & 28.2 & 29.9 &  30.5\\ \hline
    $p_6$ [dBm] & 10.8 & 15.9 & 7.8  &  10.4\\ \hline
    $p_7$ [dBm] & 12.8 & 17.1 & 13.2 &  11.5\\ \hline
    $p_8$ [dBm] & 16.3 & 17.1 & 18.0 &  16.9\\ \hline
\end{tabular}
\end{table}

\section{Applicability and discussion}
\label{sec:Applicability}

The applicability of the different methods is summarized in Table~\ref{tb:Applicability}. 

\begin{table}[ht]
\caption{Applicability of the different methods: \textsuperscript{\textdagger} requires a new training dataset; \textsuperscript{\textdaggerdbl} requires an experimental training dataset; \textsuperscript{\textasteriskcentered} requires parameter fitting of the model.}
\centering
\begin{tabular}{c||c|c|c|c}
model     & CNN & DE & CNN+DE & GD\\ \hline\hline
    Inference speed & High & Medium & Low & Low\\\hline
    Training speed & Low & Low & n.a. & n.a. \\\hline
    Scalability & \multirow{2}{2em}{ No\textsuperscript{\textdagger}} & \multirow{2}{2em}{Low} & \multirow{2}{2em}{Medium} & \multirow{2}{2em}{High} \\
    to parameters & & & &\\ \hline
    Experimental & Yes\textsuperscript{\textdaggerdbl} & Yes & Yes\textsuperscript{\textdaggerdbl} & Yes\textsuperscript{\textasteriskcentered} \\\hline

\end{tabular}
\label{tb:Applicability}
\end{table}

The CNN provides `instant' inference. The inference time is increased if higher accuracy is required by additional optimization using DE. The more general DE, and GD methods require significantly increased optimization time: in the order of minutes to hours, depending on the complexity of the scenario and computing power. On the other hand, the DE and GD do not require any pre-training and are fully flexible with dynamically changing scenarios, such as variable signal load and adding/removing a pump and/or other components in the setup. Whenever such change is required, the CNN needs a new dataset and re-training. This limitation can also be extended to physics-informed NN models, which are still partially constrained to the parameters varied during training. However, the stand-alone CNN is not constrained by the underlying physical dynamics beyond what could be learned through the training dataset. This means that its predictions, especially ones that extrapolate beyond the dataset range, are not necessarily solutions to the true model in (\ref{eq:Raman}) and may thus be non-realizable in practice, or even nonphysical. Grey-box approaches such as CNN+DE, CNN+GD, and other physics-informed models \cite{raissi2019physics} are among the methods to guide the optimization to extrapolate. Nevertheless, only applying GD or DE through a physically accurate white-box model ensures the prediction/solution follows physics laws. Physics-informed models may require additional validation of their predictions to ensure physical accuracy during inference.

The flexibility of the GD method, as well as the offline (i.e. based on a model) version of the DE, comes with the requirement that the optimization model matches to a high degree the real-life setup, which may in turn requires some parameter fitting (such as fiber parameters and connector losses). Assuming that, all methods can be applied to experimental and live network scenarios, as shown for DE and CNN+DE applied to the case under test in \cite{Soltani_OE}.

\section{Conclusions}
\label{sec:Conclusions}
Popular strategies for the optimization of optical amplifiers were summarized and compared. The example of a bi-directional Raman amplifier was studied with the target of flattening the frequency and distance gain spectrum. All studied methods achieve similar performance with approx. 1 dB of frequency-flatness at loss-compensated transmission, and approx. 3 dB of maximum power excursion when both frequency and distance flatness is the target. Each method poses its advantages and disadvantages based on the application scenario and the required speed for training, inference, and parameter fitting, as well as support for dynamically changing environments. 

\section{Acknowledgements}
This work was supported by the Villum Foundation through the VIllum Young Investigator OPTIC-AI project (grant no. VIL29334), and the Villum Investigator POPCOM project (grant no. VIL54486), by the Danish National Research Foundation through the Center of Excellence SPOC (grant no. DNRF123), and by Ministero dell’Università e della Ricerca through the PRIN 2017 project FIRST.

\bibliographystyle{elsarticle-num}
\bibliography{biblio}

\end{document}